\documentclass[aps,pre,twocolumn,amsmath,amssymb,showpacs,floatfix]{revtex4}
\usepackage{graphicx}

\begin{document}

\title{Ideal Polymers near Scale-Free Surfaces}
\author{Yosi Hammer}
\email{Email: hammeryosi@gmail.com}
\author{Yacov Kantor}
\affiliation{Raymond and Beverly Sackler School of Physics and
Astronomy, Tel Aviv University, Tel Aviv 69978, Israel}
\date{\today}

\begin{abstract}
The number of allowed configurations of a polymer is reduced
by the presence of a repulsive surface resulting in an entropic
force between them. We develop a method to calculate the entropic
force, and detailed pressure distribution, for long ideal polymers
near a scale-free repulsive surface. For infinite polymers
the monomer density is related to the electrostatic potential near
a conducting surface of a charge placed at the point where the
polymer end is held. Pressure of the polymer on the surface
is then related to the charge density distribution in the electrostatic
problem. We derive explicit expressions for pressure distributions
and monomer densities for ideal polymers near a two- or
three-dimensional wedge, and for a circular cone in three
dimensions. Pressure of the polymer diverges near sharp corners
in a manner resembling (but not identical to) the electric field
divergence near conducting surfaces. We provide formalism for
calculation of all components of the total force in situations
without axial symmetry.
\end{abstract}

\maketitle

\section{Introduction}

Statistical mechanics of long polymers near surfaces has been the
subject of numerous studies since the beginning of polymer physics
and has many important applications \cite{deGennes1979scaling}.
Problems of polymers near surfaces possess interesting relations
to critical phenomena \cite{Binder83,Eisenriegler1993}.
Current experimental methods allow manipulation and detailed study
of individual molecules revealing their conformations and properties
\cite{SMreview,leuba2001}. The atomic force microscope
\cite{binnig1986,morita2002,sarid94} (AFM) is an important
tool whose positional accuracy enables the study of the mechanical
response of single molecules to applied forces in natural conditions
and in various geometries. The spatial and force resolution of such
experiments enables measurement of relatively small deformations
of the molecules, and in that regime the interaction between the
molecule and the probes may become significant. Influence of the
shape of the probe on the elastic response of flexible polymers has
been discussed in several works
\cite{BKK_EPL88,Maghrebi2011,Maghrebi2012}, and it was shown that
several important physical properties of such systems are independent
of microscopic details of the molecule. Polymers grafted to flexible
membranes influence their shapes and physical properties
\cite{Evans97,Auth03,Guo09,Laradji02,Nikolov07,Werner10}.
Therefore, it is important to understand the detailed nature of the
interaction between polymers and surfaces.

The size of a polymer can be characterized by its root-mean-squared
end-to-end distance $R_e$. Frequently, this quantity has a simple
power law dependence on the number of monomers $N$, as $R_e=aN^\nu$
where $a$ is a microscopic length, such as monomer size, while the Flory
exponent $\nu=1/2$ for {\em ideal polymers} (IPs) that are allowed to
self-intersect in any space dimension $d$, and has $d$-dependent
values for real polymers in good solvent \cite{deGennes1979scaling}.
For a long flexible polymer containing $N$ monomers ($N\gg1)$, the
number of possible configurations is ${\cal N}\sim b^NN^{\gamma-1}$,
where $b$ is the effective model-dependent coordination number
and $\gamma$ is a universal exponent. For IPs in free
space $\gamma=\gamma_f=1$ and the power law factor in  ${\cal N}$
disappears. (For polymers in good solvents $\gamma_f$ exceeds
unity \cite{deGennes1979scaling}.)

\begin{figure}[t]
\includegraphics[width=8cm]{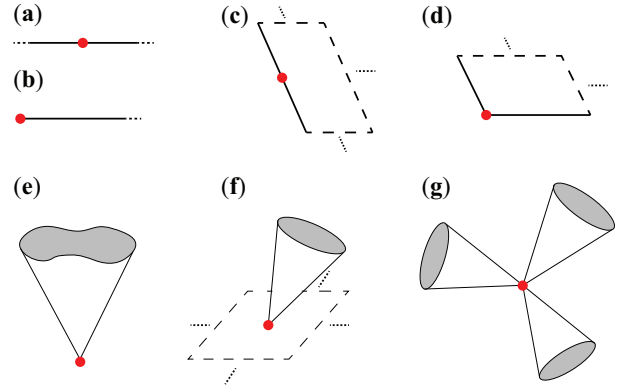} \caption{(Color online)
Scale-free surfaces and their special points (full
circles). All surfaces extend to infinity from their special points
as indicated by the dashed lines. Grey areas indicate
truncation surfaces for graphical representation, while dashed lines
represent similarly truncated surfaces. Dots indicate directions in
which the infinite objects are extended. (a) Infinite, and (b)
semi-infinite lines in $d=2$ or 3.  Semi-infinite (c) half-plane
and (d) quarter-plane in $d=3$. (e) Cone with convoluted cross section.
Complex shapes created by (f) attaching apex of a cone
to a plane, or (g) by joining apices of several cones.
}
\label{fig:SFshapes}
\end{figure}

For large $N$ there is a range of distances between $a$ and
$R_e$ where, in free space, the polymer exhibits self-similar
scale-invariant behavior. The presence of boundaries can introduce
new length scales. However,
there is a group of surfaces, called {\em scale-free} (SF),
or {\em scale-invariant}, such that geometry has no
characteristic length scale, i.e., they remain invariant under
coordinate transformation  $\mathbf{r}\to\lambda\mathbf{r}$,
when the origin of coordinates is placed at a special point.
Such surfaces as (infinite) circular cones (their apices
serve as the special points), or wedges in two and three
dimensions, will be discussed in detail in this work.
Fig.~\ref{fig:SFshapes} depicts a variety of such shapes.
Complex geometries can be made by joining special points
of SF surfaces, as demonstrated in Figs.~\ref{fig:SFshapes}f and g.
When an end-point of a polymer is attached to a special point
of repulsive SF surface, its exponent $\nu$ is not affected
but the prefactor in the relation between $R_e$ and $N$ may
change. However, SF surface {\em can} modify exponent  $\gamma$
(see Ref.~\cite{Maghrebi2012}).  E.g., $\gamma=1/2$
\cite{chandrasekhar1943} for IP attached to a repulsive plane.
For the purposes of this work it is particularly convenient to
use the universal exponent $\eta$ which is related to the decay
of density correlations. Fisher's identity \cite{cardy1996scaling}
$\gamma=(2-\eta)\nu$ relates this exponent to the ones
mentioned earlier. For IPs in free space $\eta=\eta_f=0$, and
in the presence of SF surfaces $\eta\ge0$. The total number of
configurations of IPs becomes ${\cal N}\sim b^NN^{-\eta/2}$.

\begin{figure}[t]
\includegraphics[width=5cm]{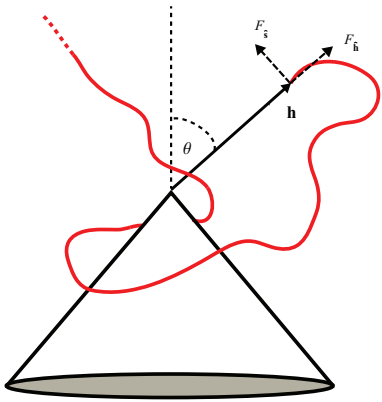} \caption{(Color online) Polymer
with one end held at position $\mathbf{h}$ from a scale-free surface.
The entropic force between the polymer and the surface, which is
also the force that acts on the point that holds the end of the
polymer has component $F_{\hat{\mathbf{h}}}$ parallel to the vector
$\mathbf{h}$, and components $F_{\hat{\mathbf{s}}}$ perpendicular to
that direction.}
\label{fig:geometry}
\end{figure}

Consider a setup, where one end of a long IP is held
at position $\mathbf{h}$ relative to the special point
of SF geometry, such as the tip of the cone in Fig.~\ref{fig:geometry}).
(The vector $\mathbf{h}$ does {\em not} have to be along some
special symmetry axis.) A significant part of the force exerted
by the polymer is coming from distances comparable with
$h\equiv|\mathbf{h}|$, while as
$N$ grows the tail of the polymer wanders-away from the surface.
For $h\ll R_e$ the total force that the polymer exerts on the
surface, or alternatively the force $\mathbf{F}$ exerted by the
surface on the polymer, becomes independent of $R_e$. In that
limit, the only dimensionally possible form for $\mathbf{F}$ is
$k_BT/h$, and therefore the component of the total force in the
direction of $\hat{\mathbf{h}}=\mathbf{h}/h$ is
\begin{equation}
F_{\hat{\mathbf{h}}}\equiv\mathbf{F}\cdot\hat{\mathbf{h}}
=\mathcal{A}\frac{k_BT}{h},\label{eq:force form}
\end{equation}
where $k_B$ is the Boltzmann constant and $T$ is the temperature.
It was demonstrated \cite{Maghrebi2011} that the dimensionless
amplitude $\mathcal{A}$  in this relation for the {\em radial}
component of the force is  independent of the  direction
$\hat{\mathbf{h}}$ and only depends on \textit{universal} exponents,
i.e., exponents which do not depend on the microscopic details of
the monomers, but depend only on a small number of parameters
such as geometry, dimensionality and the presence of self avoiding
interaction; for a IP,
\begin{equation}\label{eq:force amp}
\mathcal{A}=\eta.
\end{equation}
When $\eta_f\ne0$, as in polymers in good solvent, this
expression becomes  $\mathcal{A}=\eta-\eta_f$.
In Ref.~\cite{Maghrebi2011}, $\eta$ was found for
IPs in the cases of a cone and a wedge. In this work, we re-derive
this law and obtain the value of $\mathcal{A}$ from
different considerations, and relate it to the pressure
distribution along the boundaries.

In a non-symmetric situation, as depicted in Fig.~\ref{fig:geometry}
the force may have additional components in non-radial direction
$\hat{\mathbf{s}}\perp\hat{\mathbf{h}}$ which is given by
\begin{equation}\label{Eq:ForcePerpForm}
F_{\hat{\mathbf{s}}}\equiv\mathbf{F}\cdot\hat{\mathbf{s}}
=\mathcal{A}(\hat{\mathbf{h}},\hat{\mathbf{s}})\frac{k_BT}{h},
\end{equation}
where the value of this dimensionless amplitude depends both on the
direction of the point where the polymer is held and on
the direction of the particular force component. We provide a procedure
for calculating $\mathcal{A}(\hat{\mathbf{h}},\hat{\mathbf{s}})$, and
relate it to the pressure distribution.

In Sec.~\ref{Sec:general_formalism} we derive a general formalism
for calculation of Green and partition functions for SF surfaces,
and use it to calculate the force between the polymer and the surface.
In Sec.~\ref{sec:general_surface} we expand the formalism which has
been previously used for flat surfaces (Sec.~\ref{sec:flat})
to derive general expressions for monomer density and pressure on
a SF surface. We also demonstrate a relation between the polymer
problem and an electrostatic problem of a point charge located
near a conducting surface, and demonstrate various
symmetries of the solutions. Specific surface shapes (circular
cone in $d=3$, wedge in $d=2$ and $d=3$) are solved in
Sec.~\ref{sec:cone and wedge}.
Finally, in Sec.~\ref{sec:ring} we extend our formalism to polymers
held at both ends and to ring polymers.

\section{General formalism for confined ideal polymers}
\label{Sec:general_formalism}

\subsection{Ideal polymers confined by arbitrary shapes}

IP statistics are closely related to the statistics of
random walks (RWs) and diffusion problems \cite{wiegel1986introduction}.
An IP with $N+1$ monomers can be modeled as an $N$-step
RW. Consider a RW starting from the point $\mathbf{h}$
on a $d$-dimensional hypercubic lattice. The total number of such
walks is $\mathcal{N}_f(N)=(2d)^N$. In the presence of
confining boundaries, we denote the total number of walks starting
from $\mathbf{h}$ which {\em do not} cross the boundaries as
$\mathcal{N}_{\text{nc}}(\mathbf{h},N)$, and the number of walks
which start at $\mathbf{h}$ and end at $\mathbf{r}$
without crossing the boundaries as $\mathcal{N}_{\text{nc}}
(\mathbf{h},\mathbf{r},N)$.
In order to investigate the properties of long polymers in confined
spaces we focus on the following two functions: the (normalized)
partition function (or random walker survival probability),
\begin{equation}
Z(\mathbf{h},N)=\frac{\mathcal{N}_{\text{nc}}(\mathbf{h},N)}
{\mathcal{N}_f(N)},\label{eq:partition}
\end{equation}
and the Green function (or propagator),
\begin{equation}
G(\mathbf{h},\mathbf{r},N)=\frac{\mathcal{N}_{\text{nc}}
(\mathbf{h},\mathbf{r},N)}
{a^d\mathcal{N}_f(N)}.\label{eq:green}
\end{equation}
The division by the volume of lattice cell $a^d$ converts the probability
into probability density in the continuum description. When
the relevant distances of the problem are much larger than the
lattice constant $a$, both functions can be approximated as
continuous functions which obey the diffusion equations
\cite{wiegel1986introduction},
\begin{equation}\label{eq:diffusion partition}
\begin{split}
&\frac{\partial Z(\mathbf{h},N)}{\partial N}=D\nabla_{\mathbf{h}}^2
Z(\mathbf{h},N),\\
&Z(\mathbf{h},0)=1,
\end{split}
\end{equation}

\begin{equation}\label{eq:diffusion green}
\begin{split}
&\frac{\partial G(\mathbf{h},\mathbf{r},N)}{\partial N}=
D\nabla_{\mathbf{r}}^2G(\mathbf{h},\mathbf{r},N)\,\\
&G(\mathbf{h},\mathbf{r},0)=\delta^d(\mathbf{h}-\mathbf{r}),
\end{split}
\end{equation}
where the  diffusion constant $D=a^2/2d$, while the subscripts
$\mathbf{h},\mathbf{r}$ of the Laplacians indicate variables  with
respect to which the derivatives are taken. In order to exclude all
the walks that cross the boundaries, we require that both $Z$ and $G$
vanish on the boundaries. In that respect, a long polymer near a
{\em repulsive} wall corresponds to diffusion near an {\em absorbing}
surface.

Since either end of a RW can be considered its beginning
or end, the Green function satisfies the reciprocity relation
\cite{morse1953methods}
\begin{equation}\label{Eq:reciprocity}
G(\mathbf{r}_1,\mathbf{r}_2,N)=G(\mathbf{r}_2,\mathbf{r}_1,N).
\end{equation}

The partition and Green functions are related by
\begin{equation}
Z(\mathbf{h},N)=\int G(\mathbf{h},\mathbf{r},N) d^d\mathbf{r}.
\end{equation}
From $G$ and $Z$ we can calculate the monomer density at a point
$\mathbf{r}$ in the allowed space,
\begin{equation}\label{eq:mon density}
\rho_N(\mathbf{h},\mathbf{r})=\int_0^N G(\mathbf{h},\mathbf{r},n)
Z(\mathbf{r},N-n)dn/Z(\mathbf{h},N).
\end{equation}
The expression in the numerator decays with increasing $N$ due
to absorbing boundary condition. Since $\int
G(\mathbf{h},\mathbf{r},n)Z(\mathbf{r},N-n)d^d\mathbf{r}=Z(\mathbf{h},N)$
independently of the value of $n$, the total number of monomers is
$\int\rho_N(\mathbf{h},\mathbf{r})d^d\mathbf{r}=N$.
In most of the examples we will consider  monomer density for infinite
polymers, and therefore the total number of monomers will be infinite.
Both $G$ and $Z$ in the integrand of  Eq.~\eqref{eq:mon density}
satisfy diffusion equations with absorbing boundaries, i.e. they both
vanish at the boundaries, and approach them with finite slopes.
Therefore, the density itself vanishes quadratically close to
the boundary.

\subsection{Ideal polymers confined by scale-free shapes}

In the presence of a scale-free surface, neither
Eq.~\eqref{eq:diffusion partition}
nor the boundary surface introduce any length scale
into the problem. Therefore, the partition function depends only on the
dimensionless ratio $\mathbf{w}\equiv\mathbf{h}/\sqrt{DN}$, i.e.
$Z(\mathbf{h},N)=H(\mathbf{w})$, where $H$ is a dimensionless function.
In terms of the reduced variable Eq.~\eqref{eq:diffusion partition}
becomes \cite{Maghrebi2012}
\begin{equation}
\label{Eq:diffusion_red}
\nabla_{\mathbf w}^2 H+\frac{1}{2}{\bf w}\cdot\vec{\nabla}_{\mathbf w}H=0.
\end{equation}
When the size of the polymer is significantly larger than $h$, i.e., $w\ll 1$
the second term in the equation becomes negligible, and the equation reduces
to
\begin{equation}\label{Eq:HarmonicH}
\nabla_{\mathbf w}^2 H=0.
\end{equation}
In the presence of scale-free surfaces, it is useful to describe the
polymer in a coordinate system that separates the radial part from all other
coordinates, as is done in spherical or polar coordinates. In many of these
systems \cite{moon1971field}, the Laplace operator can be written in the form
\begin{equation}\label{eq:diffusion sep}
\nabla_{\mathbf{w}}^2=w^{1-d}\frac{\partial}{\partial w}\left(w^{d-1}
\frac{\partial}{\partial w}\right)+w^{-2}\nabla_{S_{d-1}}^2,
\end{equation}
where $\nabla_{S_{d-1}}^2$ is the Laplace-Beltrami operator acting
on the $d-1$ non-radial coordinates \cite{chavel1984eigenvalues}.
Since the boundary conditions on $H$ are independent of $w$, we expect
that for $w\ll 1$ the solution can be expressed as a product of a power of
$w$ and an angular function
$\Theta(\theta_1,\dots,\theta_{d-1})\equiv\Theta(\hat{\mathbf{w}})$,
where $\hat{\mathbf{w}}\equiv\mathbf{w}/w$.
In the limit $w\ll 1$ the large-$N$ expression becomes applicable,
and it follows that $Z\sim N^{-\eta/2}$. This means that for $w\ll 1$,
\begin{equation}\label{Eq:HorZassymp}
H(\mathbf{w})\approx w^\eta\Theta(\hat{\mathbf{w}})\ \
{\rm or}\ \ Z(\mathbf{h})\approx (h/\sqrt{DN})^\eta\Theta(\hat{\mathbf{h}})
\end{equation}
By substituting
this expression into Eq.~\eqref{Eq:HarmonicH} and using
Eq.~\eqref{eq:diffusion sep} we obtain an eigenvalue equation
\begin{equation}\label{Eq:eigenvalue eta}
\nabla_{S_{d-1}}^2\Theta=\eta(2-d-\eta)\Theta,
\end{equation}
that determines $\eta$ and the corresponding eigenfunction
$\Theta(\hat{\mathbf{w}})$. This equation has an infinite
number of eigenvalues and eigenfunctions, but, since $Z$ (or
$H$) is a positive function, we are interested only in the
``ground state" solution that is always positive, and corresponds
to the lowest value of $\eta$. For example \cite{Maghrebi2012},
in the case of a polymer in a wedge of opening angle $2\alpha$
in $d=2$, there is only one angular variable $\theta$ measured,
say, from the symmetry axis of the wedge; in this system
$\eta=\pi/2\alpha$, while $\Theta(\theta)=\cos(\pi\theta/2\alpha)$.
For a $d=3$ circular cone, of apex angle (between the symmetry axis
and the surface of the cone) $\alpha$, the value of $\eta$ is
determined \cite{Maghrebi2012} by finding the smallest degree $\eta$
of Legendre function satisfying $P_\eta(\cos\alpha)=0$. The
corresponding $\Theta(\theta,\phi)=P_\eta(\cos\theta)$, where $\theta$
is measured from the symmetry axis, and the function is independent
of $\phi$ due to symmetry of the problem. For a $d$-dimensional cone,
Eq.~\eqref{Eq:eigenvalue eta} was solved by Ben-Naim and Krapivsky
\cite{ben2010kinetics}. Their solution was used in
\cite{Maghrebi2012,Maghrebi2011} to find the force amplitude for IPs
near cones. Another example for a geometry where
Eq.~\eqref{Eq:eigenvalue eta} can be solved is a cone with elliptical
cross section (see Ref. \cite{boersma1990electromagnetic}).

From Eq.~\eqref{Eq:HorZassymp} the free energy of the polymer is
$\mathcal{F}=-k_BT(\eta\ln h+\ln\Theta(\hat{\mathbf{h}}))+{\rm const.}$,
from which the force is compiled as
\begin{equation}\label{Eq:Ah}
F_{\hat{\mathbf{h}}}=-\frac{{\partial\mathcal{F}}}{\partial r}=
\eta\frac{k_BT}{h},
\end{equation}
i.e., the amplitude that was defined in Eq.~\eqref{eq:force form},
$\mathcal{A}=\eta$ is independent of the direction of
$\hat{\mathbf{h}}$, as stated in Eq.~\eqref{eq:force amp}.
For the amplitude in one of the perpendicular directions
$\hat{\mathbf{s}}$, we need to take a similar derivative with respect to
coordinate $r_{\hat{\mathbf{s}}}$ perpendicular to $\hat{\mathbf{h}}$.
\begin{equation}\label{Eq:As}
F_{\hat{\mathbf{s}}}=-\frac{{\partial\mathcal{F}}}
{\partial r_{\hat{\mathbf{s}}}}=
\frac{\Theta^{(\hat{\mathbf{s}})}(\hat{\mathbf{h}})}
{\Theta(\hat{\mathbf{h}})}\frac{k_BT}{h},
\end{equation}
where $\Theta^{(\hat{\mathbf{s}})}$ denotes a (angular) derivative
of $\Theta$ on a unit sphere in direction of $\hat{\mathbf{s}}$,
such as $\partial/\partial\theta$ in the spherical coordinate system.
Thus the amplitude in Eq.~\eqref{Eq:ForcePerpForm} is
$\mathcal{A}(\hat{\mathbf{h}},\hat{\mathbf{s}})=
\Theta^{(\hat{\mathbf{s}})}(\hat{\mathbf{h}})/
\Theta(\hat{\mathbf{h}})$.

If the end of a polymer is tethered to the origin by a string of
length $h$, but is allowed to fluctuate in non-radial direction,
the function $\Theta(\hat{\mathbf{h}})$, that must be normalized,
is the probability density for the orientation $\hat{\mathbf{h}}$.
Since $\Theta$ is positive in the allowed space (and vanishes
only on the boundaries) it will frequently have a single maximum,
such as the position of the symmetry axis in the case of a wedge or
a cone, although multiple maxima can be created by, say, properly
shaping the cross section of a cone. This probability is
independent of temperature, and therefore the fluctuations of
the end-point will also be temperature independent. In simple
geometries the fluctuations will be ``large", i.e., occupy
most of the available directions.

The Green function has dimensions [length]$^{-d}$ and satisfies
Eq.~\eqref{eq:diffusion green}. It can  be written using the same
dimensionless variable, $\mathbf{w}$ as well as
$\mathbf{v}\equiv\mathbf{r}/\sqrt{DN}$, as
\begin{equation}\label{Eq:Gstruct}
G(\mathbf{h},\mathbf{r},N)=(DN)^{-d/2}Y(\mathbf{w},\mathbf{v}),
\end{equation}
where $Y$ is a dimensionless function.
For $DN\gg h^2$ ($w\ll 1$) the system loses its detailed dependence
on the initial condition, resulting in a function of $\mathbf{r}$
with $\mathbf{h}$ dependent prefactor. Thus, we attempt a solution
of the form
\begin{equation}\label{eq:green sep}
\tilde{G}(\mathbf{r},N)=Cg(r,N)\Theta(\hat{\mathbf{r}}),
\end{equation}
where unit vector $\hat{\mathbf{r}}\equiv\mathbf{r}/r
=\{\theta_1,\dots\theta_{d-1}\}$
describes the non radial coordinates,
and $C$ is a (dimensional) prefactor containing $h,D$. For $g(r,N)$ we use the
expression
\begin{equation}
g(r,N)=r^xN^y\exp\left(-\frac{r^2}{4DN}\right).\label{eq:scale inv ansatz}
\end{equation}
Note that the exponent $\exp\left(-\frac{r^2}{4DN}\right)$ is {\em exactly}
the same  as in the description of an IP in free space.
By using Eqs.~\eqref{eq:diffusion sep}-\eqref{eq:scale inv ansatz}
in Eq.~\eqref{eq:diffusion green} we get
\begin{equation}
\left[\frac{r^2}{DN}\left(\frac{d}{2}+x+y\right)+x(2-d-x)\right]\Theta=
\nabla_{S_{d-1}}^2\Theta .\label{eq:separation2}
\end{equation}
 In order for Eq.~\eqref{eq:separation2} to hold for arbitrary values
of $r$, the coefficient of $r$ must vanish, leading to
\begin{equation}
\frac{d}{2}+x+y=0.\label{eq:xy relation}
\end{equation}
 The value of $x$ is determined by the eigenvalue equation
\begin{equation}
\nabla_{S_{d-1}}^2\Theta=x(2-d-x)\Theta.\label{eq:eigenvalue x}
\end{equation}
This (angular) equation coincides with Eq.~\eqref{Eq:eigenvalue eta},
but, unlike $H$ in  Eq.~\eqref{Eq:HarmonicH}, the function $\tilde{G}$ that
we are seeking is {\em not} harmonic. Obviously, the value of
$x$ in this equation will coincide with $\eta$ that was found
in the calculation of $H$, as well as the function
$\Theta(\hat{\mathbf{r}})$ will be the same as
$\Theta(\hat{\mathbf{w}})$ describing $H$. (We seek the
``ground state" value of $x$ since the function $\tilde{G}$
must be positive.) We  shall henceforth substitute $\eta$
for $x$. It is shown below that such value of $x$ indeed produces
a correct description of the partition function.

 Thus we have a solution for the diffusion equation near a
scale-free surface. This solution \textit{does not} satisfy the
initial condition in Eq.~\eqref{eq:diffusion green} and does not
properly describe the statistics of short polymers, where the size
of the polymer approaches $h$. However, $\tilde{G}$ approaches
the exact solution for the Green function of long ($\sqrt{DN}\gg h$)
IPs near SF surface. Since the form of the Green function must be
described by Eq.~\eqref{Eq:Gstruct}, we must choose the constant in
Eq.~\eqref{eq:green sep} as
$C=ch^\eta/D^{\eta+d/2}$, where $c$ is a dimensionless constant
(that depends on $\hat{\mathbf{h}}$), leading to
\begin{equation}\label{eq:scale invariant solution full}
\tilde{G}=c\left(\frac{1}{\sqrt{DN}}\right)^d
\left(\frac{h}{\sqrt{DN}}\right)^\eta
\left(\frac{r}{\sqrt{DN}}\right)^\eta
{\rm e}^{-r^2/4DN}\Theta(\hat{\mathbf{r}}).
\end{equation}
Integration of this expression over the $d$-dimensional space confined
by the surfaces, leads (up to a dimensionless prefactor) to
the value of $Z(\mathbf{h},N)=H(\mathbf{w})\sim
(h/\sqrt{DN})^\eta=w^\eta$, i.e., the correct behavior of $Z$.

Note that from the definition of $G$ (Eq.~\eqref{eq:green}) and the
boundary conditions, $\Theta$ must be a positive function that vanishes
on the boundaries.

When the geometry is complicated, and analytical solution of
Eq.~\eqref{eq:eigenvalue x} cannot be obtained, the force amplitude
can be evaluated numerically. This process can be simplified by
considering the average position of the polymer end point,
\begin{equation}\label{eq:eta relation to end point}
R_{e,\tilde{G}}^2\equiv\frac{\int r^2\tilde{G}(\mathbf{r},N)d^d\mathbf{r}}
{\int\tilde{G}(\mathbf{r},N)d^d\mathbf{r}}=2DN(\eta+d).
\end{equation}
[Angular integrals in the numerator and denominator are identical
and cancel, while the radial integrals are simple products of powers and
Gaussians and lead to this result.]  Since $\tilde{G}$ approaches the
exact solution in the limit $N\rightarrow\infty$
we can write a formula for the force amplitude,
\begin{equation}
\eta=\lim_{N\rightarrow\infty}\frac{R_e^2}{2DN}-d.\label{eq:eta and end point}
\end{equation}
Note that in free space $\eta=0$, and we recover the usual mean squared
end-to-end distance for an IP/random walk $R_e^2=a^2N$.
When we confine the polymer by holding it near the boundary the mean
squared end-to-end distance grows but it is still linearly proportional
to the number of monomers. Using Eq.~\eqref{eq:eta and end point},
the force amplitude can be evaluated from numerical solution of the
diffusion equation or from simulations of random walks in confined
spaces.

\section{Ideal polymer near a plane}\label{sec:flat}

The problem of an IP near a repulsive plane was considered in
Refs.~\cite{bickel2001local,breidenich2007shape,jensen2013pressure}.
In this section we expand the approach used in \cite{bickel2001local}
to general $d$ and set the stage for the treatment of more complicated
surfaces.

In $d$ dimensions positions in half-space space are described by
$\mathbf{r}=(r_1,...,r_d)\equiv(\mathbf{R},r_\perp)$) with
$r_\perp>0$. For an IP with one end fixed at $\mathbf{h}=(0,\dots,0,h)$,
the Green function can be found using the method of images
\cite{chandrasekhar1943}:
\begin{equation}
\begin{split}G(\mathbf{h},\mathbf{r},N)=
\left(\frac{1}{4\pi DN}\right)^{d/2}\exp\left(-\frac{R^2}{4DN}\right)\times\\
\left\{ \exp\left(-\frac{(r_\perp-h)^2}{4DN}\right)-
\exp\left(-\frac{(r_\perp+h)^2}{4DN}\right)\right\}.
\end{split}
\label{eq:green flat}
\end{equation}
The corresponding partition function can be found by integrating
Eq.~\eqref{eq:green flat} over $\mathbf{r}$,
\begin{equation}\label{eq:partition flat}
Z(\mathbf{h},N)=\text{erf}\left(h/\sqrt{4\pi DN}\right).
\end{equation}

Using Eqs.~\eqref{eq:mon density}, \eqref{eq:green flat} and
\eqref{eq:partition flat}, and taking the $N\rightarrow\infty$
limit we get for $d>2$,
\begin{eqnarray}\label{eq:flat density}
& \rho(\mathbf{h},\mathbf{r})=\displaystyle\frac{1}{4D}
\frac{\Gamma\left(\frac{d}{2}-1\right)}{\pi^{d/2}}\frac{r_\perp}{h}
 \left\{ [R^2+(r_\perp-h)^2]^{1-d/2}\right.\nonumber \\
 &\left.-[R^2+(r_\perp+h)^2]^{1-d/2}\right\} .
\end{eqnarray}
(Henceforth, quantities without index $N$ will denote infinite polymer limit.)
For $d=2$,
\begin{equation}
\rho(\mathbf{h},\mathbf{r})=\frac{1}{4\pi D}\frac{r_\perp}{h}
\ln\frac{R^{2}+(h+r_\perp)^{2}}{R^{2}+(h-r_\perp)^{2}}.
\end{equation}

When a planar surface is distorted by infinitesimal amount
$\Delta(\mathbf{R})$ by shifting it from $r_\perp=0$ to
$r_\perp=\Delta(\mathbf{R})$), the resulting change in the number of
available conformations modifies the free energy of the polymer by an amount
\begin{equation}
\Delta\mathcal{F}_N=\int\limits _{r_\perp=0}
\left[P_N(\mathbf{R})\Delta(\mathbf{R})\right]d^{d-1}\mathbf{R}
+O(\Delta(\mathbf{R})^2),\label{eq:free energy change flat}
\end{equation}
where $P_N(\mathbf{R})$ is the entropic pressure of the polymer
on the surface at position $\mathbf{R}$. Thus, the pressure represents
a variational derivative of the free energy, and for a polymer with
one end held at $\mathbf{h}$ it can be written in terms of the Green
function \cite{bickel2001local} as
\begin{equation}\label{eq:general pressure}
P_N(\mathbf{h},\mathbf{R})=\frac{k_BTD}{Z(\mathbf{h},N)}\int_0^N
\frac{\partial G(\mathbf{r},\mathbf{h},n)}{\partial r_\perp}
\frac{\partial Z(\mathbf{r},N-n)}{\partial r_\perp}dn,
\end{equation}
where $\mathbf{r}=(\mathbf{R},r_\perp)$, and the derivatives are
evaluated at $r_\perp=0$. Equation \eqref{eq:mon density} can be used
to rewrite this expression via
the monomer density $\rho(\mathbf{r})$ ,
\begin{equation}\label{eq:pressure density relation flat}
P_N(\mathbf{h},\mathbf{R})=\frac{Dk_BT}{2}\frac{\partial^2}
{\partial r_\perp^2}\rho_N(\mathbf{h},\mathbf{r}).
\end{equation}
From Eqs.~\eqref{eq:flat density}, \eqref{eq:pressure density relation flat}
we find the polymer pressure on the plane in the limit $N\rightarrow\infty$,
\begin{equation}
P(R)=\frac{\Gamma(d/2)}{\pi^{d/2}}\frac{k_BT}{(R^2+h^2)^{d/2}}.
\label{eq:flat pressure}
\end{equation}

It should be noted that the infinite-$N$ expressions for the
density and the pressure apply to finite-$N$ situations when
$DN\gg r^2, h^2$. For smaller $N$ these quantities cannot be
expressed in such simple terms. If a polymer is confined to a
finite volume and both its ends are free to move, a different 
approach needs to be used to calculate the pressure distribution 
(see, e.g., Ref.~\cite{Grosberg72}).

\section{Ideal Polymers near General Scale-Free Surfaces}
\label{sec:general_surface}

Equation \eqref{eq:mon density} provides a general expression
for calculation of monomer density of IP for arbitrary
confining surfaces. Usually, such $\rho_N$ will be a very
complicated function.  We will demonstrate that for
scale-free surfaces for sufficiently large $N$ the
expressions for density (and also for pressure) approach
an $N$-independent form that is significantly simpler than
the small-$N$ expressions.

\subsection{Monomer Density for Infinite Polymers}
\label{subsec:simplify_integral}

Calculation of monomer density $\rho_N(\mathbf{h},\mathbf{r})$ in
Eq.~\eqref{eq:mon density} requires integration of the product
$G(\mathbf{h},\mathbf{r},n)Z(\mathbf{r},N-n)$ over $n$ varying from
0 to $N$. In free space the Green function
$G(\mathbf{h},\mathbf{r},n)$ is very small for $n$ such that
$\sqrt{Dn}\ll |\mathbf{h}-\mathbf{r}|$), because random walk from
$\mathbf{h}$ is ``too short"  to reach $\mathbf{r}$.
Similarly, for $\sqrt{Dn}\gg |\mathbf{h}-\mathbf{r}|$ the
walk is ``too long" to be at $\mathbf{r}$ with a significant
probability. Thus, $G$ in free space peaks when $\sqrt{Dn}$
is of order of $|\mathbf{h}-\mathbf{r}|$. In the presence
of absorbing boundaries, the large $n$ decay is even stronger.
In the presence of scale-free surfaces, for long polymers
($DN\gg h^2, r^2$) it is possible to divide the integral
$\int_0^N$ in Eq.~\eqref{eq:mon density} into
$\int_0^{n_1}+\int_{n_1}^N$, where $r^2,h^2\ll Dn_1\ll DN$,
and show that for fixed $x_1=n_1/N$, in the limit $N\to\infty$
the second integral divided by $Z(\mathbf{h},N)$ vanishes. This
feature is quantitatively demonstrated in Appendix
\ref{app:integral split}. Thus, only the first integral includes
significant contributions to the density. In its range ($n<n_1\ll N$)
we can assume $Z(\mathbf{r},N-n)\approx Z(\mathbf{r},N)$
and take it out of the integration so that in the infinite-$N$ limit
the density is
$\rho(\mathbf{h},\mathbf{r})=\lim_{N\rightarrow\infty}
\left[Z(\mathbf{r},N)/Z(\mathbf{h},N)\right]
\int_0^\infty G(\mathbf{h},\mathbf{r},n)dn$, or using
Eq.~\eqref{Eq:HorZassymp},
\begin{equation}\label{eq:simplification}
\rho(\mathbf{h},\mathbf{r})=
\frac{\Theta(\hat{\mathbf{r}})}{\Theta(\hat{\mathbf{h}})}
\left(\frac{r}{h}\right)^\eta\int_0^\infty
G(\mathbf{h},\mathbf{r},n)dn.
\end{equation}
This simplification enables us to perform the integral and derive
analytical expressions for the monomer density.

Since the density $\rho$ in Eq.~\eqref{eq:simplification} depends only
on the integral of $G$ it is convenient to define
\begin{equation}
\Phi(\mathbf{h},\mathbf{r})\equiv\int_0^\infty G(\mathbf{h},\mathbf{r},n)dn.
\end{equation}
From Eq.~\eqref{eq:diffusion green},
\begin{eqnarray}\label{Eq:Poisson}
\nabla_{\mathbf{r}}^2\Phi(\mathbf{h},\mathbf{r})&=&\int_0^\infty
\nabla_{\mathbf{r}}^2G(\mathbf{h},\mathbf{r},n)dn\nonumber\\
&=& (1/D)(G(\mathbf{h},\mathbf{r},\infty)-G(\mathbf{h},\mathbf{r},0))
\nonumber\\
&=&-(1/D)\delta^d(\mathbf{h}-\mathbf{r}),
\end{eqnarray}
i.e., the density at $\mathbf{r}$ is related to the potential of a point
charge at $\mathbf{h}$
\begin{equation}\label{Eq:rho_Phi}
\rho(\mathbf{h},\mathbf{r})=
\frac{\Theta(\hat{\mathbf{r}})}{\Theta(\hat{\mathbf{h}})}
\left(\frac{r}{h}\right)^\eta\Phi(\mathbf{h},\mathbf{r}).
\end{equation}

\subsection{Some properties of monomer density and pressure}

The expression for calculation of monomer density of an infinite
IP in Eq.~\eqref{eq:simplification}, requires knowledge of the
exact Green function or electrostatic potential. These are frequently
expressed as an infinite sum of functions. Since the density
$\rho(\mathbf{h},\mathbf{r})$
is singular for $\mathbf{h}=\mathbf{r}$ such expansions of density
do not always converge. Even the simple expression for pressure
on flat surfaces in Eq.~\eqref{eq:pressure density relation flat}
can be expanded in powers of $r/h$, but will converge only for
$r<h$. Alternatively, it can be expanded in the powers
of $h/r$, and will converge only for $h<r$. This situation
will recur for more complicated surfaces discussed
in the following section.

If the expression for monomer density in an infinite polymer
\eqref{eq:simplification} is combined with the reciprocity relation
of the Green function in Eq.~\eqref{Eq:reciprocity}, or Eq.~\eqref{Eq:rho_Phi}
is combined with the reciprocity property of electrostatic potential,
we can relate the monomer densities in the situation when the polymer
starting point $\mathbf{r}_1$ and the observation point $\mathbf{r}_2$
interchange their roles
\begin{equation}\label{Eq:rho_reciprocity}
\rho(\mathbf{r}_1,\mathbf{r}_2)=
\left[\frac{\Theta(\hat{\mathbf{r}}_2)}{\Theta(\hat{\mathbf{r}}_1)}\right]^2
\left(\frac{r_2}{r_1}\right)^{2\eta}\rho(\mathbf{r}_2,\mathbf{r}_1).
\end{equation}
From Eq.~\eqref{Eq:Gstruct} it follows that in scale-free geometries
$G(\lambda\mathbf{r}_1,\lambda\mathbf{r}_2,\lambda^2n)=\lambda^{-d}
G(\mathbf{r}_1,\mathbf{r}_2,n)$, which can be used with
Eq.~\eqref{eq:simplification} to obtain
\begin{equation}\label{Eq:expansion}
\rho(\lambda\mathbf{r}_1,\lambda\mathbf{r}_2)=\lambda^{2-d}
\rho(\mathbf{r}_1,\mathbf{r}_2).
\end{equation}
(This relation can also be obtained from the properties of $\Phi$ under
rescaling.) This means that the structure of the density function
can be slightly simplified: If instead of variables $\mathbf{r}_1$
and $\mathbf{r}_2$ we use the direction of the two vectors and their
lengths $r_1$ and $r_2$, when the ratio of the lengths is $x=r_2/r_1$,
then by choosing $\lambda=1/r_1$ in Eq.~\eqref{Eq:expansion} we find
that $\rho(\mathbf{r}_1,\mathbf{r}_2)=r_1^{2-d}
f(x,\hat{\mathbf{r}}_1, \hat{\mathbf{r}}_2)$. We therefore expect that
the calculation of density function will involve an expansion of the
solution in the dimensionless ratio $x$.

Let us now consider a Kelvin transform of coordinates where the new position
is obtained by inverting the old position with respect to a sphere of radius
$h$: $\mathbf{r}_2=(h/r_1)^2\mathbf{r}_1$ (Under this transformation
$\mathbf{h}$ maps into itself.) If the potential
$\Phi_1(\mathbf{h},\mathbf{r}_1)$ of the original
problem  is known, then \cite{doob2001classical}
$\Phi_2(\mathbf{h},\mathbf{r}_2)
\equiv(r_1/h)^{d-2}\Phi_1(\mathbf{h},\mathbf{r}_1)$
also solves Eq.~\eqref{Eq:Poisson}. Usually, performing Kelvin
transform requires similar
transformation of the boundary surfaces, but in this case the boundary
conditions are independent of the length $r_1$ and therefore are automatically
satisfied for $r_2$. This relation together with Eq.~\eqref{Eq:rho_Phi}
leads to the conclusion that
\begin{equation}\label{Eq:rho_Kelvin}
\rho(\mathbf{h},\mathbf{r}_2)=(r_1/h)^{d-2-2\eta}\rho(\mathbf{h},\mathbf{r}_1).
\end{equation}
This feature conveniently connects the values of the density for, say, $r_1/h=y<1$
with the values of density at $r_2/h=1/y>1$, i.e., the density for $r>h$ can be
reconstructed from the density at $r<h$. For $r_1\ll h$ the electrostatic
potential $\Phi\sim\Theta(\hat{\mathbf{r}}_1)r_1^\eta/Dh^{d-2}$, since in
that region it satisfies the same equation as $Z$. Therefore, from
Eq.~\eqref{Eq:rho_Phi} we find in that limit
$\rho(\mathbf{h},\mathbf{r}_1)\approx
A[\Theta(\hat{\mathbf{r}}_1)]^2r_1^\eta/Dh^{d-2+2\eta}$,
where $A$ is a dimensionless constant. By using Eq.~\eqref{Eq:rho_Kelvin}
we can now determine that for $r_2=h^2/r_1\gg h$ the density is
$\rho(\mathbf{h},\mathbf{r}_2)
\approx A[\Theta(\hat{\mathbf{r}}_2)]^2/Dr_2^{d-2}$
with {\em the same} coefficient $A$. The latter relation does not
depend on $h$, as could be expected in that region. Since $G$ is a
solution of diffusion equation, the expression must include the prefactor
$1/D$ of dimension [length]$^{-2}$. (The same conclusion follows for
Eqs.~\eqref{Eq:Poisson} and \eqref{Eq:rho_Phi}.) Therefore,
aside from angular term, the result is the only dimensionally
possible expression for the density.

The method presented in Sec.~\ref{sec:flat} to compute the entropic
pressure of the polymer in half-space can be generalized to any
regular surface (i.e., any surface that appears flat when observed
from an infinitesimal distance). For a general surface, we define
the distortion $\Delta(\tilde{\mathbf{r}})$ to be in the direction
perpendicular to the surface, where $\tilde{\mathbf{r}}$ is a point
on the surface. The derivative with respect to $r_\perp$ in
Eqs.~\eqref{eq:flat pressure} now represents
derivative in the direction locally perpendicular to the surface.

Pressure on the boundary corresponds to the second derivative with respect to
coordinate perpendicular to the boundary. If $\mathbf{r}_2$ and $\mathbf{r}_1$
are related by Kelvin transform, as mentioned above, and are on the boundary
of the surface, then from Eq.~\eqref{Eq:rho_Kelvin} it follows that pressures
at corresponding points are related by
\begin{equation}
P(\mathbf{h},\mathbf{r}_2)=(r_1/h)^{d+2-2\eta}P(\mathbf{h},\mathbf{r}_1).
\end{equation}
From these relations, by repeating the argument analogous to the one in the
previous paragraph, or directly from the expressions of $\rho$ at very large
and very small distances, we can establish
that for $r_2\gg h$ the expression for pressure has the $h$-independent
dimensionally correct form
$P(\mathbf{h},\mathbf{r}_2)\approx
B[\Theta^{(\hat{\mathbf{s}})}(\hat{\mathbf{r}}_2)]^2k_BT/r_2^d$,
where $\Theta^{(\hat{\mathbf{s}})}(\hat{\mathbf{r}}_2)$ is the derivative of
$\Theta$ on the unit sphere in direction $\hat{\mathbf{s}}$ perpendicular
to the boundary, evaluated on the boundary,
and $B$ is some dimensionless constant. Using the
arguments outlined above we conclude that  at short distances $r_1\ll h$,
the pressure becomes
\begin{equation}\label{Eq:P_power}
P(\mathbf{h},\mathbf{r}_1)\approx B[\Theta^{(\hat{\mathbf{s}})}
(\hat{\mathbf{r}}_1)]^2
k_BT r_1^{2(\eta-1)}/h^{d+2(\eta-1)},
\end{equation}
with the same $B$.

\subsection{Pressure and the total force}
From Eqs.~\eqref{eq:pressure density relation flat} and \eqref{Eq:rho_Phi}
the expression for the pressure at a point on a surface can be written
as
\begin{eqnarray}
P(\mathbf{h},\mathbf{r})&=&\frac{Dk_BT}{2}\frac{\partial^2}{\partial r_\perp^2}
\left[\frac{\Theta(\hat{\mathbf{r}})}{\Theta(\hat{\mathbf{h}})}
\left(\frac{r}{h}\right)^\eta\Phi(\mathbf{h},\mathbf{r})\right]\nonumber\\
&=&Dk_BT \nabla_{\mathbf{r}}
\left[\frac{\Theta(\hat{\mathbf{r}})}
{\Theta(\hat{\mathbf{h}})}\left(\frac{r}{h}\right)^\eta\right]
\cdot\nabla_{\mathbf{r}}\Phi(\mathbf{h},\mathbf{r}),
\end{eqnarray}
where we used the fact that both functions vanish on the boundary
and their gradients are parallel to each other and perpendicular
to the boundaries. This expression can be used to
calculate the total force acting in the direction of $\hat{\mathbf{h}}$
by integrating the projection of the force on the desired direction on
the entire surface,
\begin{eqnarray}
&F_{\hat{\mathbf{h}}}&=\int_S d\mathbf{S}\cdot
\hat{\mathbf{h}}P(\mathbf{h},\mathbf{r})\\
&=&\!\!  Dk_BT\!\! \int \nabla_{\mathbf{r}}\!\!\cdot\!\!\left\{
\hat{\mathbf{h}}\nabla_{\mathbf{r}}
\left[\frac{\Theta(\hat{\mathbf{r}})}
{\Theta(\hat{\mathbf{h}})}\left(\frac{r}{h}\right)^\eta\right]
\!\! \cdot\!\! \nabla_{\mathbf{r}}\Phi(\mathbf{h},\mathbf{r})
\right\}d^d\mathbf{r}.\nonumber
\end{eqnarray}
By applying the divergence operator and using the fact that the
function in the first square brackets is harmonic while the
electrostatic potential satisfies Eq.~\eqref{Eq:Poisson} we find that,
\begin{eqnarray}\label{Eq:FfromPparallel}
F_{\hat{\mathbf{h}}}&=&Dk_BT \int\hat{\mathbf{h}}\cdot
\nabla_{\mathbf{r}}
\left[\frac{\Theta(\hat{\mathbf{r}})}
{\Theta(\hat{\mathbf{h}})}\left(\frac{r}{h}\right)^\eta\right](1/D)
\delta^d(\mathbf{h}-\mathbf{r})d^d\mathbf{r}\nonumber\\
&=&k_BT \int\frac{\partial}{\partial r}
\left[\frac{\Theta(\hat{\mathbf{r}})}
{\Theta(\hat{\mathbf{h}})}\left(\frac{r}{h}\right)^\eta\right]
\delta^d(\mathbf{h}-\mathbf{r})d^d\mathbf{r}\nonumber\\
&=&\eta k_BT/h\ ,
\end{eqnarray}
which coincides with the general expression for the force
in Eq.~\eqref{eq:force form} with $\mathcal{A}=\eta$,
as was seen directly in Eq.~\eqref{Eq:Ah}.

Similar calculations can be performed for the force components in
an arbitrary direction. If we choose some direction
$\hat{\mathbf{s}}\perp\hat{\mathbf{h}}$, we can repeat the above
calculation with the new projection direction $\hat{\mathbf{s}}$.
Now on the first line of Eq.~\eqref{Eq:FfromPparallel} we will
have the product $\hat{\mathbf{h}}\cdot\nabla_{\mathbf{r}}$ replaced
by $\hat{\mathbf{s}}\cdot\nabla_{\mathbf{r}}$, which will result
in the derivative acting only on $\Theta(\hat{\mathbf{r}})$ in direction
$\hat{\mathbf{s}}$ (such as $(1/r)(\partial/\partial\theta)$ in
three-dimensional spherical coordinates) leading to exactly
the same result as in Eq.~\eqref{Eq:As}.

\section{Specific Geometries}
\label{sec:cone and wedge}

We will now discuss the monomer density and the entropic pressure of an IP
on a wedge in $d=2$ and $3$ a cone in $d=3$. The Green functions for
the cone and wedge geometries can be found in Appendix
\ref{ap:Green functions}. The calculation was
performed according to the procedure described in section
\ref{sec:general_surface}. For simplicity throughout
this section we consider cases where the end of the polymer
is held along the symmetry axis of the cone or wedge.

\begin{figure}
\includegraphics[width=4.5cm]{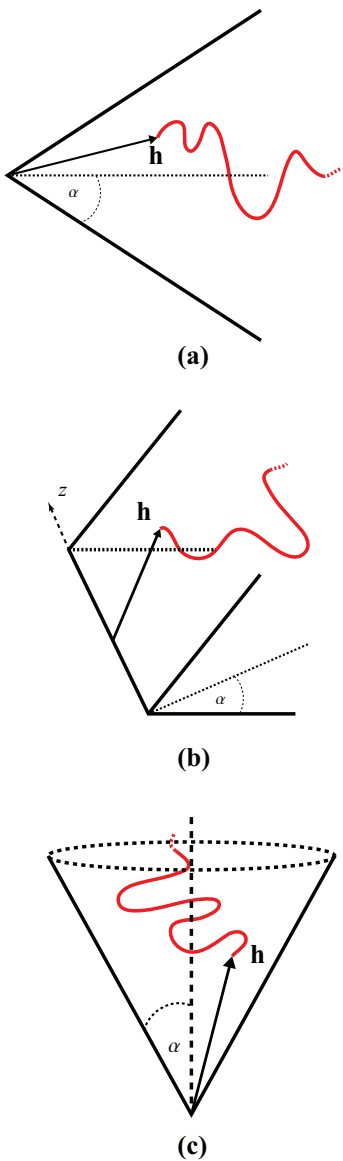}
\caption{(Color online) An ideal polymer confined to scale-free
spaces: (a) wedge in $d=2$, (b) wedge in $d=3$, and (c) circular
cone in $d=3$. }
\label{fig:geometry wedges and cone}
\end{figure}

\subsection{Wedge in $d=2$}

\begin{figure}[t]
\includegraphics[width=8cm]{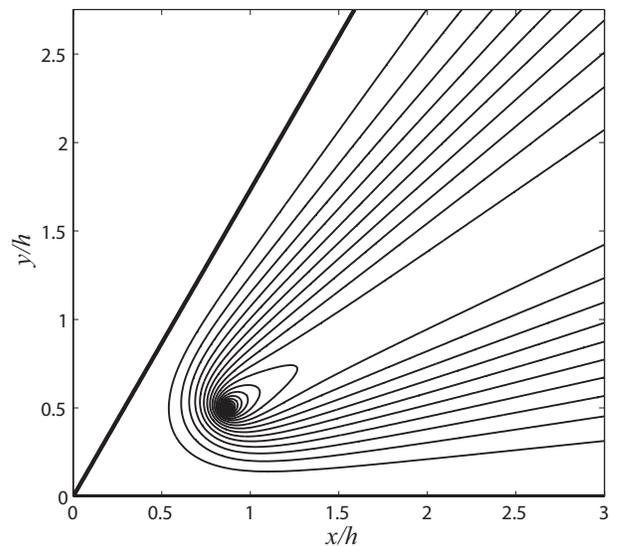}
\caption{Contour plot of the monomer density for a long ideal
polymer with one end held at a distance $h$ from the tip of a wedge
($d=2$) with opening angle $2\alpha=\pi/3$. The boundary is marked
by the thick solid line. The density is highest at the origin of the
polymer, and the constant density lines are equally separated on a
linear scale.}
\label{fig:density}
\end{figure}

\begin{figure}[t]
\includegraphics[width=7cm]{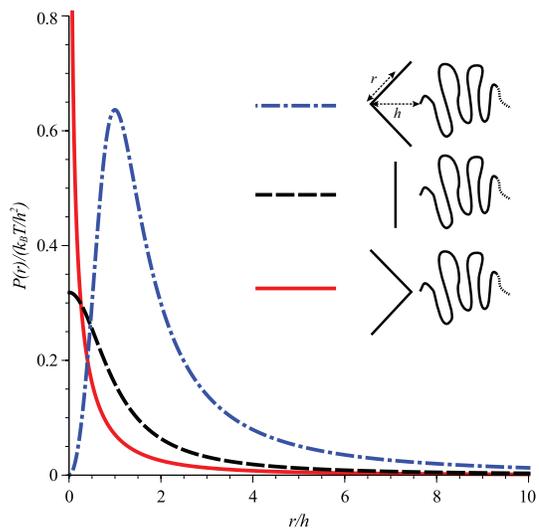}
\caption{(Color online) Scaled entropic pressure of a long ideal
polymer on a wedge ($d=2$) as a function of the scaled distance from
the tip for three opening angles of the wedge, $\alpha=\pi/4$ (dot-dashed
line), $\alpha=\pi$ (dashed line), $\alpha=3\pi/4$ (solid line).}
\label{fig:pressure plot}
\end{figure}

Consider a wedge defined in polar coordinates by $-\alpha<\theta<\alpha$
(Fig.~\ref{fig:geometry wedges and cone}a). One end of an IP
is held at a distance $h$ from the corner, along the symmetry axis
of the wedge. The Green function for this geometry is given in Appendix
\ref{ap:Green functions} (Eq.~\eqref{eq:green function wedge}). In the
range where $h^2\ll DN$ and $r^2\lesssim DN$, the sum in
Eq.~\eqref{eq:green function wedge} becomes a power series and the lowest
power dominates. Thus the Green function converges to the general
form presented in Eq.~\eqref{eq:scale invariant solution full}. The
force amplitude is the lowest power in the series \eqref{eq:wedge roots}, i.e.,
\begin{equation}
\eta=\pi/2\alpha.\label{eq:wedge force amp}
\end{equation}
 In order to derive the monomer density in the wedge, we follow the
procedure outlined in section~\ref{sec:general_surface}. For
$N\rightarrow\infty$ we get
\begin{equation}
\begin{split}\rho(\mathbf{r})=\frac{1}{2\pi D}
\cos\left(\frac{\pi\theta}{2\alpha}\right)
\left(\frac{r}{h}\right)^{\pi/2\alpha}\times\\
\tanh^{-1}\frac{2\cos(\pi\theta/2\alpha)}
{(h/r)^{\pi/2\alpha}+(r/h)^{\pi/2\alpha}}\;.
\end{split}
\label{eq:mon density wedge}
\end{equation}
 The monomer density in a wedge with $\alpha=\pi/3$ is depicted in
Fig.~\ref{fig:density}. The derivative perpendicular to the surface
in this geometry is $\frac{\partial}{\partial r_\perp}=\frac{1}{r}
\frac{\partial}{\partial\theta}$.
Using Eqs.~\eqref{eq:pressure density relation flat} and
\eqref{eq:mon density wedge}
we find the entropic pressure on the surface of the wedge (still for
$N\rightarrow\infty$),
\begin{equation}
P(r)=\frac{\pi}{4\alpha^2}\frac{k_BT}{r^2}
\frac{1}{1+(h/r)^{\pi/\alpha}}\;.\label{eq:pressure wedge}
\end{equation}
 It is interesting to note the asymptotic behavior of the pressure
for small $r$,
\begin{equation}
\lim_{r\rightarrow0}P(r)\propto r^{\pi/\alpha-2}\rightarrow\begin{cases}
0 & 0<\alpha<\pi/2\\
\mathrm{const.} & \alpha=\pi/2\\
\infty & \pi/2<\alpha<\pi
\end{cases}.\label{eq:asymp wedge}
\end{equation}
 When the polymer is held outside the wedge ($\alpha>\pi/2$) the pressure
on the tip diverges. This behavior can be seen In
Fig.~\ref{fig:pressure plot}, where we plot the pressure on the wedge
for three  different opening angles. The singularity at the tip of the
wedge is similar (but not identical) to the one found in the electric
field near the tip of a charged conductor. The analogy to electric
fields is not surprising, since we have seen that the monomer density
is related to the electrostatic potential of a point charge
(Eq.~\eqref{Eq:rho_Phi}). The electric field near the tip of a
conducting wedge scales
as $r^{\pi/2\alpha-1}$ \cite{jackson}, whereas the polymer pressure
scales as $r^{\pi/\alpha-2}$ (see Eq.~\eqref{eq:asymp wedge}). For
a flat plane ($\alpha=\pi/2$), both powers vanish.

\subsection{Wedge in $d=3$}

The boundary of a wedge in  3-dimensional space is defined in cylindrical
coordinates by $-\alpha<\theta<\alpha$
(Fig.~\ref{fig:geometry wedges and cone}b).
Consider a case where one end of the polymer is held at a distance $h$ from
the corner, along the symmetry axis of the wedge, at $z=0$. The Green function
for this geometry is given in Eq.~\eqref{eq:wedge green function d=3}. It is
obtained by multiplying the Green function of the wedge in $d=2$ by the
$d=1$ free space propagator. When $z^2\ll DN$, the Green function is
independent of $z$ and when in addition $h^2\ll DN $, $r^2\lesssim DN $,
it assumes the general form of Eq.~\eqref{eq:scale invariant solution full}.
The force amplitude is the one found in $d=2$
(Eq.~\eqref{eq:wedge force amp}). The monomer density in the wedge in the
limit $N\rightarrow\infty$ is
\begin{multline}
\rho(r,\theta,z)=\frac{1}{\pi^{1/2}D\alpha}
\frac{\left(r/h\right)^{\pi/2\alpha}}
{(r^2+h^2+z^2)^{1/2}}\cos\left(\frac{\pi\theta}{2\alpha}\right)\times\\
\sum_{i=1}^\infty\Gamma\left(
\frac{1}{2}+\frac{\pi i}{2\alpha}\right)\sin\left(\frac{\pi i}{2}\right)
\cos\left(\frac{i\pi\theta}{2\alpha}\right)\times\\
\left(\frac{rh}{r^2+h^2+z^2}\right)^{i\pi/2\alpha}\times\\
_2\tilde{F}_1\left[\frac{1}{4}+\frac{i\pi}{4\alpha},\frac{3}{4}+
\frac{i\pi}{4\alpha},1+\frac{i\pi}{2\alpha},
\left(\frac{2rh}{r^2+h^2+z^2}\right)^2\right],
\label{eq:monomer_density_3d_wedge}
\end{multline}
 where $_2\tilde{F}_1$ is the regularized hypergeometric function.
The pressure on the surface of the wedge is
\begin{multline}
P(r,z)=\frac{k_BT\pi^{3/2}}{8\alpha^3}\frac{(r/h)^{\pi/2\alpha}}
{r^2(r^2+h^2+z^2)^{1/2}}\times\\
\sum_{i=1}^\infty i\sin\left(\frac{\pi i}{2}\right)
\left(\frac{rh}{r^2+h^2+z^2}\right)^{\pi i/2\alpha}
\Gamma\left(\frac{1}{2}+\frac{i\pi}{2\alpha}\right)\times\\
_2\tilde{F}_1\left[\frac{1}{4}+\frac{i\pi}{4\alpha},\frac{3}{4}+
\frac{i\pi}{4\alpha},1+\frac{i\pi}{2\alpha},
\left(\frac{2rh}{r^2+h^2+z^2}\right)^2\right].\label{eq:pressure 3d wedge}
\end{multline}
Note that at the tip of the wedge we get the same irregular behavior
that was encountered for $d=2$, i.e. for $r\ll h$,
\[
\underset{r\rightarrow0}{\lim}P\propto\frac{r^{\pi/\alpha-2}}
{(h^2+z^2)^{(\pi/\alpha+1)/2}}.
\]
The unusual influence of the geometry of two- and three-dimensional 
wedges is  known in the theory or critical phenomena and the remarkable
effects of such geometries have been studied in detail \cite{Hanke99a}
in the context of critical adsorption of liquids.

\subsection{Circular Cone in $d=3$}

Now consider a cone defined in spherical coordinates by $\theta<\alpha$
(Fig.~\ref{fig:geometry wedges and cone}c). The polymer is confined to the
cone with one end held at a distance $h$ from the tip along the symmetry
axis of the cone. The Green function for this geometry is given in
Appendix \ref{ap:Green functions}. Once more, it contains a sum which
becomes a power series when $h^2\ll DN$, $r^2\lesssim DN$.
As in the case of the wedge, in the limit $N\rightarrow\infty$ the
first term in the series is dominant, and the Green function converges
to the general form of Eq.~\eqref{eq:scale invariant solution full}.
The force amplitude, $\eta$, is the lowest root of the equation
\begin{equation}
P_\eta(\cos\alpha)=0,\label{eq:cone force amp}
\end{equation}
 where $P_\eta$ is the Legendre function. When we apply the procedure
described above to calculate the monomer density
in the limit $N\rightarrow\infty$, we find
\begin{equation}
\begin{split} & \rho(\mathbf{r})=-\frac{1}{2\pi D\sqrt{hr}}P_\eta(\mu)
\left(\frac{r}{h}\right)^\eta\times\\
 & \sum_{i=1}^\infty\left(\left[(1-\mu_0^2)\frac{\partial}
 {\partial\mu}P_{\eta_{i}}(\mu_{0})\frac{\partial}{\partial\eta_i}
 P_{\eta_i}(\mu_0)\right]^{-1}\times\right.\\
 & \left.P_{\eta_i}(\mu)\begin{cases}
(r/h)^{\eta_i+1/2} & r<h\\
(h/r)^{\eta_i+1/2} & r>h
\end{cases}\right),
\end{split}
\label{eq:mon density cone}
\end{equation}
where $\eta_i$ are the roots of Eq.~\eqref{eq:cone force amp},
in ascending order, $\eta=\eta_1$, $\mu=\cos\theta$ and $\mu_0=\cos\alpha$. The
derivative in the direction perpendicular to the surface is the same
as in the case of the wedge (with $\theta$ being the polar angle
in spherical coordinates). The pressure on the surface is
\begin{equation}
\begin{split} & P(r)=-\frac{k_BT}{2\pi r^3}\frac{\partial}
{\partial\mu}P_\eta(\mu_0)\times\\
 & \begin{cases}
(\frac{r}{h})^{\eta+1}\sum_{i=1}^\infty\frac{(r/h)^{\eta_i}}
{\frac{\partial}{\partial\eta_i}P_{\eta_i}(\mu_0)} & r<h\\
(\frac{r}{h})^\eta\sum_{i=1}^\infty\frac{(h/r)^{\eta_i}}
{\frac{\partial}{\partial\eta_i}P_{\eta_i}(\mu_0)} & r>h
\end{cases}
\end{split}
,\label{eq:pressure cone}
\end{equation}
 Note that for $r\rightarrow0$ we get $P\propto r^{2(\eta-1)}$.
The singular asymptotic behavior of the pressure on the tip of the
cone is similar to the one found on the wedge (Eq.~\eqref{eq:asymp wedge}).
It is also similar to the behavior of electric fields near the tip
of a conducting cone, where the field scales as $r^{\eta-1}$. For
a flat plane ($\alpha=\pi/2$, $\eta=1$), the powers vanish. In fact,
if we consider a point charge $q$ held at height $h$ above a grounded
conducting plane, the electric field on the plane is identical with
the polymer pressure on a flat plane if we replace $k_BT/2\pi$ by
$2qh$.

\section{Polymers Held at both Ends\protect \\
and Ring Polymers} \label{sec:ring}

\begin{figure}[t]
\includegraphics[width=5cm]{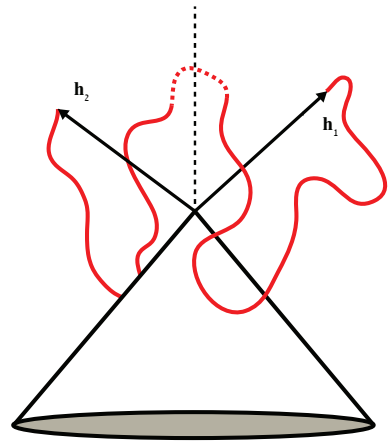} \caption{(Color online) Ideal polymer
with both ends held near a scale-free surface.}
\label{fig:ring}
\end{figure}

Consider an IP held at {\em both} ends at points $\mathbf{h}_1$ and
$\mathbf{h}_2$  close to scale-free surface as depicted in
Fig.~\ref{fig:ring}. The monomer density at a point
$\mathbf{r}$ in the allowed space is
\begin{equation}
\rho_N(\mathbf{h}_1,\mathbf{h}_2,\mathbf{r})=
\frac{\int_0^NG(\mathbf{h}_1,\mathbf{r},n)G(\mathbf{r},\mathbf{h}_2,N-n)dn}
{G(\mathbf{h}_1,\mathbf{h}_2,N)}.\label{eq:density both ends}
\end{equation}
For large $N$, such that $h_1^2,h_2^2\ll DN$, we can follow the same
reasoning as in the subsection \ref{subsec:simplify_integral}, and
divide the integral $\int_0^N$ into {\em three} parts
$\int_0^{n_1}+\int_{n_1}^{N-n_1}+\int_{N-n_1}^N$, where $n_1$ was
chosen such that $h_1^2,h_2^2\ll Dn_1\ll DN$. (Note, that when
the integration variable $n$ is replaced by $N-n$ in the third
integral, it becomes similar to the first integral with the roles
of $\mathbf{h}_1$ and $\mathbf{h}_2$ reversed.) The first Green function
in the integrand of Eq.~\eqref{eq:density both ends} provides a
significant contribution
to the integral $\int_0^{n_1}$, while the second Green function is
similarly significant in $\int_{N-n_1}^N$. Both functions are negligible
in $\int_{n_1}^{N-n_1}$. In fact it can be shown that for fixed $x_1=n_1/N$,
in the $N\to\infty$ limit, the latter integral divided by
$G(\mathbf{h}_1,\mathbf{h}_2,N)$ vanishes. In the range of the first
integral ($n<n_1\ll N$) we can assume
$G(\mathbf{r},\mathbf{h}_2,N-n)\approx G(\mathbf{r},\mathbf{h}_2,N)$
and take it out of the integration so that for large $N$ the
contribution of this integral to the density becomes
$\left[G(\mathbf{r},\mathbf{h}_2,N)/G(\mathbf{h}_1,\mathbf{h}_2,N)\right]
\int_0^{x_1N}G(\mathbf{h}_1,\mathbf{r},n)dn$. For large $N$ the
ratio of the Green functions preceding the integral can be replaced
by the ratio of $\tilde{G}$ functions defined in Eq.~\eqref{eq:green sep},
which, by using Eq.~\eqref{eq:scale invariant solution full}
and the reciprocity relation \eqref{Eq:reciprocity}, becomes
$(r/h_1)^\eta [\Theta(\hat{\mathbf{r}})/\Theta(\hat{\mathbf{h}}_1)]$.
Similar treatment, with the roles of $\mathbf{h}_1$
and $\mathbf{h}_2$ reversed, can be performed for the integral
$\int_{(1-x_1)N}^N$. The results discussed above become exact in
the $N\to\infty$ limit leading to
\begin{eqnarray}\label{eq:both ends dens calculation}
\rho(\mathbf{h}_1,\mathbf{h}_2,\mathbf{r})&=&
\left(\frac{r}{h_1}\right)^\eta\frac{\Theta(\hat{\mathbf{r}})}
{\Theta(\hat{\mathbf{h}}_1)}
\int\limits _0^\infty G(\mathbf{h}_1,\mathbf{r},n)dn\nonumber\\
&+&\left(\frac{r}{h_2}\right)^\eta
\frac{\Theta(\hat{\mathbf{r}})}{\Theta(\hat{\mathbf{h}}_2)}
\int\limits _0^\infty G(\mathbf{h}_2,\mathbf{r},n)dn.
\end{eqnarray}

By comparing this result with Eqs.~\eqref{eq:both ends dens calculation}
and \eqref{eq:simplification}, we see that the contribution from each
end of the chain will be equal to the density calculated before for
a polymer with one free end, i.e.,
\begin{equation}\label{eq:density additivity}
\rho(\mathbf{h}_1,\mathbf{h}_2,\mathbf{r})=
\rho(\mathbf{h}_1,\mathbf{r})+\rho(\mathbf{h}_2,\mathbf{r}),
\end{equation}
where $\rho(\mathbf{h}_i,\mathbf{r})$ is the density which
was calculated in sections \ref{sec:flat}-\ref{sec:cone and wedge}
with $\mathbf{h}=\mathbf{h}_i$. This result could be expected
since the strands leaving the end-points $\mathbf{h}_1$ and
$\mathbf{h}_2$ do not interact with each other, and the
mid-section of the polymer is so far away, that the fact
that this is a single polymer rather than two independent
strands does not influence the density. From the relation
between the monomer density and the pressure on the surface
(Eq.~\eqref{eq:pressure density relation flat}) we see that
this additive correspondence between a pair of IP strands and
a single polymer held by both ends will also apply to the pressure
and the total force on the surface.
This result can also be immediately applied to the density of
infinite ring  polymers:
\begin{equation}\label{eq:density ring}
\rho_{\text{ring}}(\mathbf{h},\mathbf{r})=
\rho(\mathbf{h},\mathbf{h},\mathbf{r})=2\rho(\mathbf{h},\mathbf{r}),
\end{equation}
as well as to the pressure exerted by such polymers.

\section{Discussion and conclusions}

In Ref.~\cite{Maghrebi2011} it was shown that the force between a polymer
and a scale-free surface can be written in terms of universal
exponents, which depend on the geometry of the surface and the nature
of the interactions between the monomers, but are independent of
the microscopic details of the system. In this paper we have shown
that the universal exponent $\eta$ also plays an important role
in the monomer density and the pressure of IPs on scale-free surfaces.
Additional calculations were needed to completely describe the pressure
and the density, but $\eta$ controlled the behavior at very short
distances. We found the general form of the Green function $\tilde{G}$
(Eq.~\eqref{eq:scale invariant solution full}) for
long IPs. By using the simple connection between the exponent
$\eta$ and the mean end-to-end distance of the polymer $R_e$
(Eq.~\eqref{eq:eta and end point}), one can measure $\eta$ by
solving the diffusion equation numerically or extracting $R_e$ from simulations.

In section~\ref{sec:general_surface} we showed that the monomer density
and  the entropic pressure can be derived from the electrostatic potential
of a point charge in a confined space. It was also shown that the density
possesses some powerful scaling properties that enable one to map the
density and the pressure from points near the origin to points far away,
and vice versa. The relation to electrostatics also
enabled the use of a formalism resembling Gauss' law in electrostatics,
to relate the total force to the pressure distribution.

Our calculations were limited to IPs. While they provide some
guidance to understanding polymers in good solvents, several
important differences exist. The presence of repulsion between
monomers modifies both exponent $\nu$ and $\eta$. The basic
expression \eqref{eq:mon density} cannot be used in its simplest
form to calculate the density because the probability of a
polymer reaching point $\mathbf{r}$ in $n$ steps is influenced
by the presence of the remaining $N-n$ steps, and proper
adjustments need to be made. Even in free space the distribution
of the the end-to-end distance of self-avoiding polymers is
significantly more complicated (see, \cite{Caracciolo00}
and references therein) than the Green function of ideal
polymers. Thus, we cannot expect such simple behaviors as
exhibited by $\tilde{G}$ in
 \eqref{eq:scale invariant solution full}. Nevertheless, we
may expect some qualitative similarities between IP and self-avoiding
polymers. Polymer adsorption to curved surfaces \cite{Eisenriegler96}
introduces yet another dimension into the problem deserving
a detailed study.

In good solvents the density of the monomers no longer decays
quadratically with the distance $x$ from the boundary. Scaling
analysis shows \cite{joanny1979effects} that close to the walls
the density scales as $x^{1/\nu}$. (In good solvent $1/\nu<2$
\cite{deGennes1979scaling}.) 
Bickel {\em et al.}~\cite{bickel2001local} used this scaling law to
compare the behavior of IPs to self-avoiding polymers near flat 
surfaces, and found numerous qualitative (and even quantitative)
parallels between the two cases. It remains to be seen if such
parallels can be found in connection with the properties discussed
in this paper. Hanke {\em et al.} \cite{Hanke99} found interesting 
depletion effects of polymers in good solvent near curved surfaces. 

The properties of the monomer density discussed in
Sec.~\ref{sec:ring} indicate that the results in this paper can
be applied to entropic
systems where both ends of the polymer are attached to scale-free
surfaces. This may provide a pathway to dealing with polymers attached
by both ends to different surfaces, such as a polymer with one
end grafted to an AFM tip and the other to a flat substrate. In
good solvents expressions like  \eqref{eq:density additivity} and
\eqref{eq:density ring} are obviously incorrect. However, like in IPs,
we expect that for very long polymer the behavior of two ends of a
polymer will be the same as that of two (interacting) polymers
with their remote ends completely free.

\begin{acknowledgments}
We thank M. Kardar for useful discussions and for comments on the manuscript.
This work was supported by the Israel Science Foundation grant 186/13.
\end{acknowledgments}

\appendix

\section{Green and Partition Functions}
\label{ap:Green functions}
Below we list the exact solutions of Eq.~\eqref{eq:diffusion green}
that were used to derive the results in this paper. These solutions
were taken from Ref.~\cite{carslaw1959conduction}.

{\em Wedge in $d=2$}: The wedge defined in
Fig.~\ref{fig:geometry wedges and cone} is described in polar coordinates
by $-\alpha<\theta<\alpha$. The exact solution
of  Eq.~\eqref{eq:diffusion green} is \cite{carslaw1959conduction}
\begin{widetext}
\begin{equation}
G(r',\theta',r,\theta,N)=\frac{1}{2\alpha DN}\exp\left(-\frac{r^2+r'^2}
{4DN}\right)
\sum_{i=1}^\infty I_{\eta_i}\left(\frac{rr'}{2DN}\right)
\cos(\eta_i\theta)\cos(\eta_i\theta'),
\label{eq:green function wedge}
\end{equation}
 where
\begin{equation}\label{eq:wedge roots}
\eta_i=\frac{i\pi}{2\alpha},\;\; i=1,2,3,\dots
\end{equation}
and $I_{\eta_i}$ is the modified Bessel function of the first
kind. The partition function can be found by integrating
Eq.~\eqref{eq:green function wedge}:
\begin{eqnarray}
Z(r',\theta',N)&=&\int G(r',\theta',r,\theta,N)rdrd\theta\nonumber\\
&=&\frac{r'}{\sqrt{\pi DN}}\exp\left(-\frac{r^2}{8DN}\right)
\sum_{i=1}^\infty \frac{1}{i}\left[I_{\frac{\eta_{i}-1}{2}}
\left(\frac{r'^2}{8DN}\right)
+I_{\frac{\eta_{i}+1}{2}}\left(\frac{r'^2}{8DN}\right)\right]
\sin^2\left(\frac{\pi i}{2}\right)\cos\left(\eta_{i}\theta'\right).
\label{eq:partition wedge}
\end{eqnarray}

{\em Wedge in $d=3$}: The solution in $d=3$ is obtained by multiplying
the Green function of a wedge in $d=2$ by the free space propagator in
$d=1$ leading to
\begin{equation}
G(r',\theta',z',r,\theta,z,N)=
\frac{1}{4\alpha\pi^{1/2}(DN)^{3/2}}\exp\left(-\frac{(z-z')^2}{4DN}\right)
\sum_{i=1}^\infty\exp\left(-\frac{r^2+r'^2}{4DN}\right)I_{\eta_{i}}
\left(\frac{rr'}{2DN}\right)\cos(\eta_{i}\theta')\cos(\eta_{i}\theta).
\label{eq:wedge green function d=3}
\end{equation}
Integrating Eq.~\eqref{eq:wedge green function d=3} one can immediately see
that the partition function for the wedge in $d=3$ does not depend
on the $z$ coordinate. In fact it is identical with the partition function
for the wedge in $d=2$ in Eq.~\eqref{eq:partition wedge}.

{\em Circular cone}: Circular cone in $d=3$ is defined in spherical coordinates
by $\theta<\alpha$ (Fig.~\ref{fig:geometry wedges and cone}c). The solution to
Eq.~\eqref{eq:diffusion green} in
the cone is \cite{carslaw1959conduction}
\begin{multline}
G(r',\theta',\phi',r,\theta,\phi,N)=-\frac{1}{4\pi DN\sqrt{rr'}}
\exp\left(-\frac{r^2+r'^2}{4DN}\right)
\sum_{i=1}^\infty\left\{ I_{\eta_i+1/2}\left(\frac{rr'}{2DN}
\right)(2\eta_i+1)\right.\times\\
 \sum_{m=0}^\infty(2-\delta_{m,0})P_{\eta_{i}}^{-m}(\mu)P_{\eta_i}^{-m}(\mu')
 \cos(m(\phi-\phi'))
 \left.\left[(1-\mu_0)^2\frac{\partial}{\partial\mu}P_{\eta_i}^{-m}(\mu_0)
 \frac{\partial}{\partial\eta}P_{\eta_i}^{-m}(\mu_0)\right]^{-1}\right\} ,
\label{eq:green function cone}
\end{multline}
 where $P_{\eta}^{-m}$ are associated Legendre functions, $\mu=\cos\theta$,
$\mu_{0}=\cos\alpha$ and ${\eta_{i}}$ are the roots of the equation
$P_{\eta}(\mu_{0})=0$
in ascending order. The Green function is somewhat simpler when the starting
point of the polymer is along the symmetry axis of the cone, i.e., $\theta'=0$.
In this case the solution does not depend on the azimuthal angle $\phi'$.
Denoting $r'=h$, we get
\begin{equation}
G(h,r,\theta,N)=-\frac{1}{4\pi DN\sqrt{hr}}
\exp\left(-\frac{r^2+h^2}{4DN}\right)
\sum_{i=1}^\infty \frac{I_{\eta_{i}+1/2}\left(\frac{rh}{2DN}\right)
(2\eta_{i}+1)P_{\eta_{i}}(\mu)}
{(1-\mu_{0}^2)\frac{\partial}{\partial\mu}P_{\eta_{i}}(\mu_{0})
\frac{\partial}{\partial\eta}P_{\eta_{i}}(\mu_{0})}.
\label{eq:green function cone 2}
\end{equation}
The partition function can be found for any $\mathbf{r}'$ by integrating
Eq.~\eqref{eq:green function cone},
\begin{eqnarray}\label{eq:partition function cone}
&Z&(r',\theta',N)=\int r^2\sin\theta G(r',\theta',\phi',r,\theta,\phi,N)
drd\theta d\phi\nonumber\\
&=&e^{-\frac{r'^2}{4DN}}\sum_{i=1}^\infty\Gamma
\left(\frac{3+\eta_i}{2}\right){}_{1}\tilde{F}_{1}
\left(\frac{3+\eta_i}{2},\frac{3}{2}+\eta_i,\frac{r'^2}{4DN}\right)
P_{\eta_i}(\mu)\left(\frac{r'^2}{4DN}\right)^{\eta_i/2}
\frac{P_{\eta_i+1}(\mu_0)-P_{\eta_i-1}(\mu_0)}{(1-\mu_0)^2
\frac{\partial}{\partial\mu}P_{\eta_i}(\mu_0)
\frac{\partial}{\partial\eta}P_{\eta_i}(\mu_0)},\
\end{eqnarray}
where $_{1}\tilde{F}_{1}$ is the regularized confluent hypergeometric
function \cite{tableRyzhik2000}. Note that due to the cylindrical
symmetry $Z$ does not depend on the azimuthal angle $\phi'$.

\section{Monomer Density in the  $N\rightarrow\infty$ limit}
\label{app:integral split}
In the calculation of the monomer density and the entropic pressure
we examined separately the contribution of different parts of the
polymer. The monomer density in the limit $N\rightarrow\infty$
is given by
\begin{equation}
\rho(\mathbf{h},\mathbf{r})=\lim_{N\rightarrow\infty}
\int\limits _{n=0}^{N}G(\mathbf{h},\mathbf{r},n)Z(\mathbf{r},N-n)dn/
Z(\mathbf{h},N).\label{eq:density1 appendix}
\end{equation}
In order to evaluate this integral we select the ratio $x_1=n_1/N$ such
that $0<x_1\ll1$ and split the integral
in Eq.~\eqref{eq:density1 appendix},
\begin{eqnarray}
\rho(\mathbf{h},\mathbf{r}) & = & \lim_{N\rightarrow\infty}
\left\{ \left(\int\limits _{0}^{n_1}+\int\limits _{n_1}^{N-n_1}
+\int\limits _{N-n_1}^{N}\right)G(\mathbf{h},\mathbf{r},n)Z(\mathbf{r},N-n)dn/
Z(\mathbf{h},N)\right\}
\label{eq:density split appendix}\\
 & = & \lim_{N\rightarrow\infty}\left\{
 N\left(\int\limits_0^{x_1}+\int\limits _{x_1}^{1-x_1}+
 \int\limits _{1-x_1}^{1}\right)G(\mathbf{h},\mathbf{r},n)Z(\mathbf{r},N-n)dx/
 Z(\mathbf{h},N)\right\} ,\nonumber
\end{eqnarray}
where we have changed the variable of integration to $x=n/N$. Using
the scaling properties of the functions $G$ and $Z$ for large $N$
($G$ scales as in Eq.~\eqref{eq:scale invariant solution full} and
$Z(\mathbf{h},N)=\left(h/\sqrt{DN}\right)^{\eta}\Theta(\hat{\mathbf{h})}$
as in Eq.~\eqref{Eq:HorZassymp}), we see that in the limit
$N\rightarrow\infty$, for $n_1<n<N-n_1$,
\begin{eqnarray*}
G(\mathbf{h},\mathbf{r},n) & \propto & \left(\frac{1}{N}
\right)^{d/2+\eta}G^*(\mathbf{h},\mathbf{r},x),\\
Z(\mathbf{r},N-n) & \propto & \left(\frac{1}{N}\right)^{\eta/2}
Z^*(\mathbf{r},1-x),
\end{eqnarray*}
where $Z^*$ and $G^*$ do not depend on $N$. Therefore,
\[
N\int\limits _{x_1}^{1-x_1}G(\mathbf{h},\mathbf{r},n)Z(\mathbf{r},N-n)dx/
Z(\mathbf{h},N)\propto\left(\frac{1}{N}\right)^{d/2+\eta-1}\rightarrow0.
\]
Since for $x\rightarrow1$ the partition function
$Z(\mathbf{r},N-n)\rightarrow1$
and becomes independent of $N$, and it is always smaller than one
(it is the survival probability of a random walker), in the same limit,
\[
N\int\limits _{1-x_1}^{1}G(\mathbf{h},\mathbf{r},n)Z(\mathbf{r},N-n)dx/
Z(\mathbf{h},N)\propto\left(\frac{1}{N}\right)^{d/2+\eta/2-1}\rightarrow0.
\]
Only the first part of the polymer will contribute to the monomer density
in the limit of an infinitely long polymer, as described in the main text,
and will lead to density in Eq.~\eqref{eq:simplification}.

\end{widetext}

\bibliographystyle{apsrev}
\bibliography{HKArxiv}
\end{document}